\documentclass[pra,twocolumn,showpacs,preprintnumbers,amsmath,amssymb]{revtex4}

\usepackage{graphicx}
\usepackage{dcolumn}
\usepackage{bm}
\newcommand{\bra}[1]{\langle#1|}
\newcommand{\ave}[1]{\langle#1\rangle}
\newcommand{\modul}[1]{|#1|}
\newcommand{\scal}[2]{\langle#1|#2\rangle}\newcommand{\ket}[1]{|#1\rangle}

\begin{document}

\preprint{APS/123-QED}

\title{Bell's inequality violation for entangled generalized Bernoulli states\\ in two spatially separate cavities}

\author{R. Lo Franco}
 \email{rosario.lofranco@alice.it}
\author{G. Compagno}
 \email{compagno@fisica.unipa.it}
  \author{A. Messina}
  \email{messina@fisica.unipa.it}
 \author{A. Napoli}
 \email{napoli@fisica.unipa.it}
\affiliation{%
INFM, MIUR and Dipartimento di Scienze Fisiche ed Astronomiche,
Universit\'a di Palermo, via Archirafi 36, 90123 Palermo, Italy }

\date{\today}

\begin{abstract}
We consider the entanglement of orthogonal generalized Bernoulli
states in two separate single-mode high-$Q$ cavities. The
expectation values and the correlations of the electric field in
the cavities are obtained. We then define, in each cavity, a
dichotomic operator expressible in terms of the field states which
can be, in principle, experimentally measured by a probe atom that
``reads'' the field. Using the quantum correlations of couples of
these operators, we construct a Bell's inequality which is shown
to be violated for a wide range of the degree of entanglement and
which can be tested in a simple way. Thus the cavity fields
directly show quantum non-local properties. A scheme is also
sketched to generate entangled orthogonal generalized Bernoulli
states in the two separate cavities.
\end{abstract}


\maketitle

\section{\label{intro}Introduction}
Quantum entanglement of spatially separate systems is at the
origin of non-local behavior \cite{epr,mand,pre}. Bell's
inequalities \cite{bell,clau} show in fact that, if the degree of
entanglement is large enough, quantum correlations are
incompatible with the ones due to the so-called local hidden
variable theories. Experimental evidence supports quantum theory,
as it was first shown by \citet{asp,asp1} using entangled
travelling photons and more recently by \citet{moeh} using an
hybrid atom-photon entanglement.

In the context of cavity quantum electrodynamics (CQED) several
schemes, based on typical atom-cavity interactions, have been
proposed for the production, in two separate single-mode cavities,
of entangled number states of the kind
$a\ket{1_11_2}+b\ket{0_10_2}$ \cite{bergou}, or
$a\ket{1_10_2}+b\ket{0_11_2}$ \cite{mey,brow,mes,compagno1}, where
$0_j,1_j$ ($j=1,2$) indicates the number of photons in the
cavity~$j$ and $\modul{a}^2+\modul{b}^2=1$. In spite of this,
there are few proposals to test Bell's inequalities for such
entangled cavity fields and an experimental test has not yet been
made. One of these proposals exploits an indirect method, the
entanglement to be brought to light being transferred to probe
atoms, for which Bell's inequality is tested in terms of atomic
pseudo-spin operators \cite{gerry}. In this case one finds that
atomic non-locality stands from the initial non-locality of the
cavity photon system. Another proposal consists in a direct
non-locality test for entangled cavity fields \cite{kim} in which
Bell's inequality to be tested is formed with parity field
operators \cite{bana1,bana2}. Within this test, classical driving
fields are coupled with the two cavities where a nonlocal state
has been already prepared. A couple of independent two-level atoms
is then sent through the cavities and each atom off-resonantly
interacts with the respective cavity field. After going out of the
cavities, the atomic states are conditionally measured and Bell's
inequality can be tested by the joint probabilities of the
two-level atoms being both in their excited or ground state.

The aim of this paper is to propose a direct test of Bell's
inequality for entangled fields in two spatially separate
single-mode cavities using appropriate measurable cavity field
operators simply implementable by the usual resonant atom-cavity
interactions. It is of particular relevance to aim at getting
entanglement between electromagnetic field states having
mesoscopic characteristics, since in such a condition the
classical-quantum border may be investigated. To this end, as well
as in the context of our procedure, it is strategic to start up
with entangled two-cavities electromagnetic fields exhibiting a
non-zero mean field in each cavity and resulting from the passage
of few resonant two-level atoms through both cavities. The reason
is that, for such states, one expects to find out correlations
between the non-zero fields of the two cavities. Quantum states of
the electromagnetic field which satisfy these requirements are,
for example, the binomial states \cite{sto,mou}. In this paper we
consider the entangled state of two separate single-mode cavities
both filled with a ``generalized Bernoulli state'' (GBS), which is
a particular case of a one-excitation generalized binomial state
\cite{sto}, briefly discussing a possible way to realize it
experimentally. We study the expectation values and the
correlations of the electric field for this entangled two-cavities
state, finding, as expected, non-zero values. In such a condition,
it is useful to define, in each cavity, a dichotomic operator
(eigenvalues $\pm1$) expressible in terms of the cavity field
states and experimentally measurable, in principle, with the help
of a resonant probe atom. We then construct a Bell's inequality
involving the correlations get established between these
dichotomic operators, discussing wether and how it may be violated
by the entanglement injected into the two-cavities system. We
moreover suggest a simple test of this Bell's inequality
violation.

This paper is organized as follows: in Sec.\ref{entsez} we
introduce the entanglement of GBSs in two separate cavities; in
Sec.\ref{correlec} we study the expectation values and the
correlations of the electric field for this entangled state; in
Sec.\ref{bellviol} we introduce the dichotomic cavity operator by
which we prove the violation of Bell's inequality, suggesting a
simple experimental test, too; in Sec.\ref{gensch} we suggest a
way to generate entangled GBSs in two separate high-$Q$
single-mode cavities and we briefly discuss the potential
experimental errors involved in this scheme; in Sec.\ref{conc} we
summarize our conclusions.

\section{\label{entsez}Entanglement of orthogonal generalized Bernoulli states}
The single-mode binomial state of the electromagnetic field was
introduced by \citet{sto} and its principal properties are
reported in literature \cite{sto,vid}. Here we are interested in
the particular ``generalized binomial state'' \cite{sto} where two
consecutive number states have the same relative phase $\phi$,
defined as
\begin{equation}
\ket{N,p,\phi}\equiv\sum_{n=0}^N\left[{N\choose
n}p^{n}(1-p)^{N-n}\right]^{1/2}e^{in\phi}\ket{n},\label{bindef}
\end{equation}
where $N$ is the maximum number of photons of the field,
$p\in]0,1[$ is the probability of a single photon occurrence,
$\phi$ is the mean phase \cite{vid} and ${N\choose
n}=N!/[(N-n)!n!]$ is the Newton binomial. This state is clearly
normalized. Some useful properties of the binomial state of
Eq.~(\ref{bindef}) are given in App.\ref{binapp}, where in
particular we prove the orthogonality condition for two binomial
states with the same $N$. More in detail, we find that two
binomial states $\ket{N,p,\phi}$ and $\ket{N,1-p,\pi+\phi}$
satisfy the condition given in Eq.~(\ref{ortog}) and they are
orthogonal. In the particular case $N=1$, the generalized binomial
state $\ket{1,p,\phi}\equiv\ket{p,\phi}$ is called ``generalized
Bernoulli state'' (GBS) \cite{sto}.

Let us now suppose that two identical separate single-mode
cavities, namely 1 and 2, are prepared in an entangled state of
the form
\begin{eqnarray}
\ket{\Psi}&=&\mathcal{N_{\eta}}\big[\ket{p_1,\vartheta_1}_1\ket{1-p_2,\pi+\vartheta_2}_2\nonumber\\
&+&\eta\ket{1-p_1,\pi+\vartheta_1}_1\ket{p_2,\vartheta_2}_2\big],\label{Entbinom1}
\end{eqnarray}
where $\eta$ is real, ${\cal N}_\eta$ is a normalization constant
and the state $\ket{p_j,\vartheta_j}_j$ ($j=1,2$) indicates that
the cavity~$j$ is in a GBS with probability of a single photon
occurrence $p_j$ and mean phase $\vartheta_j$. The GBSs relating
to the same cavity of Eq.~(\ref{Entbinom1}) satisfy the
orthogonality condition given in Eq.~(\ref{ortog}), so the state
$\ket{\Psi}$ of Eq.~(\ref{Entbinom1}) represents an entanglement
of orthogonal GBSs in two spatially separate cavities and the
normalization constant is
\begin{equation}
{\cal N}_\eta=1/\sqrt{1+\modul{\eta}^2}.
\end{equation}

Using the property given in Eq.~(\ref{binlim}), we observe that,
for limit values of $p_1,p_2$, the entangled state of
Eq.~(\ref{Entbinom1}) becomes
\begin{subequations}
\label{Ent01}
\begin{eqnarray}
\ket{\Psi}_{p_1=p_2=1}={\cal N_\eta}[\ket{1_10_2}+\eta
e^{i(\vartheta_2-\vartheta_1)}\ket{0_11_2}]\\
\ket{\Psi}_{p_1=1,p_2=0}={\cal N_\eta}[\ket{1_11_2}-\eta
e^{-i(\vartheta_1+\vartheta_2)}\ket{0_10_2}]
\end{eqnarray}
\end{subequations}
that is the entanglement of GBSs reduces to the entanglement of
number states $\ket{0},\ket{1}$.

In the following, we shall consider the entangled orthogonal GBSs
of Eq.~(\ref{Entbinom1}) as the injected state in the two cavities
and we shall study, in this state, expectation values and
correlations of the electric field and Bell's inequality
violations.

\section{\label{correlec}Expectation values and correlations of the electric field}
In this section we proceed to calculate both the expectation value
of the electric field in the GBS of a single cavity and the
correlations of the electric fields in the entangled GBSs of the
two cavities defined in Eq.~(\ref{Entbinom1}). We analyze the
dependance of the correlations on the values of the system
variables. Since we are considering single-mode electromagnetic
fields of frequency $\omega$ inside cavities of volume $V$, the
quantized electric field inside each cavity~$j$ ($j=1,2$), at the
time $t_j=0$, can be written as
$\hat{\bm{E}}_j(z_j)=\bm{\epsilon}_j\hat{E}_j(z_j)$, where
\cite{vid}
\begin{equation}
\hat{E}_j(z_j)=\sqrt{\frac{4\pi\hbar\omega}{V}}(a_j+a_j^\dag)\sin(kz_j).\label{elec}
\end{equation}
To obtain simple quantitative results, from now on we shall
consider the electric field defined in Eq.~(\ref{elec}) at the
time $t_j=0$ in the center of the cavity, where $\sin(kz_j)=1$.

We first calculate the matrix elements in the cavity~$j$ of
$\hat{E}_j$ in the basis of the two orthogonal GBSs
$\ket{p_j,\vartheta_j}$ (state~1) and
$\ket{1-p_j,\pi+\vartheta_j}$ (state~2). Using Eqs.~(\ref{elec})
and (\ref{bin1}), we obtain
\begin{eqnarray}
\ave{\hat{E}_j}_{11}&=&-\ave{\hat{E}_j}_{22}=2\sqrt{\frac{4\pi\hbar\omega p_j(1-p_j)}{V}}\cos\vartheta_j\nonumber\\
\ave{\hat{E}_j}_{12}&=&\sqrt{\frac{4\pi\hbar\omega}{V}}[(2p_j-1)\cos\varphi_j-i\sin\vartheta_j].\label{elemE}
\end{eqnarray}
The diagonal matrix element $\ave{\hat{E}_j}_{11}$ represents the
expectation value of the electric field in the GBS
$\ket{p_j,\vartheta_j}$ and in general it differs from zero.

Using Eq.~(\ref{elemE}), we now calculate the expectation value of
the electric field in each cavity~$j$ for the entangled GBSs of
Eq.~(\ref{Entbinom1}), given by
$\ave{\hat{E}_j}_{\Psi}\equiv\bra{\Psi}\hat{E}_j\ket{\Psi}$. At
the time $t_j=0$ and in the center of the cavity, we find
\begin{equation}
\ave{\hat{E}_j}_{\Psi}=2(-1)^{j-1}\sqrt{\frac{4\pi\hbar\omega
p_j(1-p_j)}{V}}\frac{1-\modul{\eta}^2}{1+\modul{\eta}^2}\cos\vartheta_j.\label{AverE}
\end{equation}
It is immediate to note that, when the entanglement is
non-maximal, that is for $\modul{\eta}\neq1$, and when
$p_1,p_2\neq0,1$ and $\vartheta_1,\vartheta_2\neq\pm\pi/2$, the
expectation value of the electric field of Eq.~(\ref{AverE})
differs from zero. On the other hand, for $p_1,p_2=0,1$, i.e. when
the entanglement is between number states, as given in
Eqs.~(\ref{Ent01}), we find that the expectation value of the
electric field in each cavity is always equal to zero for any
value of $\eta$. From Eq.~(\ref{AverE}) we also obtain that, if
$\modul{\eta}=1$, i.e. if the entanglement of
Eq.~(\ref{Entbinom1}) is maximal, the mean electric field in each
cavity vanishes, as it is expected. The correlation function of
the electric fields in the two cavities is given by
$\ave{\hat{E}_1\hat{E}_2}_{\Psi}$, and using
Eqs.~(\ref{Entbinom1}) and (\ref{elec}) we obtain
\begin{eqnarray}
\ave{\hat{E}_1\hat{E}_2}_{\Psi}=\frac{8\pi\hbar\omega}{V}
\bigg\{\frac{\eta}{1+\modul{\eta}^2}\big[f(p_1,p_2)\cos\vartheta_1\cos\vartheta_2\nonumber\\
+\sin\vartheta_1\sin\vartheta_2\big]-h(p_1,p_2)\cos\vartheta_1\cos\vartheta_2\bigg\}\label{CorrE}
\end{eqnarray}
where we have set
\begin{subequations}
\begin{eqnarray}
f(p_1,p_2)&\equiv&(2p_1-1)(2p_2-1)\label{f}\\
h(p_1,p_2)&\equiv&2\sqrt{p_1p_2(1-p_1)(1-p_2)}\label{h}.
\end{eqnarray}
\end{subequations}
A quantitative indication of the electric field correlations is
given by the \emph{covariance}
\begin{equation}
\mathcal{C}(\hat{E}_1,\hat{E}_2)=\ave{\hat{E}_1\hat{E}_2}_{\Psi}-\ave{\hat{E}_1}_{\Psi}\ave{\hat{E}_2}_{\Psi}.\label{Ecov}
\end{equation}
From Eqs.~(\ref{AverE}) and (\ref{CorrE}), we find that the
covariance for the entangled state of Eq.~(\ref{Entbinom1}) is
\begin{eqnarray}
\mathcal{C}(\hat{E}_1,\hat{E}_2)&=&
\frac{8\pi\hbar\omega}{V}\bigg\{\frac{\eta}{1+\modul{\eta}^2}\big[f(p_1,p_2)\cos\vartheta_1\cos\vartheta_2\nonumber\\
&+&\sin\vartheta_1\sin\vartheta_2\big]-\left[1-\left(\frac{1-\modul{\eta}^2}{1+\modul{\eta}^2}\right)^2\right]\nonumber\\
&&\times
h(p_1,p_2)\cos\vartheta_1\cos\vartheta_2\bigg\}.\label{Ecov1}
\end{eqnarray}
${\cal C}(\hat{E}_1,\hat{E}_2)$ is in general different from zero,
and it vanishes when $\eta=0,\pm\infty$, i.e. when the entangled
state $\ket{\Psi}$ of Eq.~(\ref{Entbinom1}) becomes a simple
product of two uncorrelated GBSs.

We now consider the covariance of Eq.~(\ref{Ecov1}) for some
particular values of the system variables. For $\eta=\pm1$ and
$p_1=p_2=1/2$ we have
\begin{equation}
\mathcal{C}_{\eta=\pm1;p_1=p_2=1/2}(\hat{E}_1,\hat{E}_2)=-\frac{4\pi\hbar\omega}{V}\cos(\vartheta_1\pm\vartheta_2)\label{Ecov2}
\end{equation}
which show the importance of the values of the phase angles
relations $\vartheta_1\pm\vartheta_2$. In fact, if
$\vartheta_1\pm\vartheta_2=\pi/2$, the covariance of
Eq.~(\ref{Ecov2}) vanishes; if instead
$\vartheta_1\pm\vartheta_2=0$, the covariance of Eq.~(\ref{Ecov2})
would be equal to the maximum value $-4\pi\hbar\omega/V$.

The electric field covariance for maximally entangled number
states of the form given in Eqs.~(\ref{Ent01}) is obtained by
setting $\eta=\pm1$, with $p_1=p_2=1$ or $p_1=1,p_2=0$ in
Eq.~(\ref{Ecov1}). Then we find
\begin{subequations}
\label{Ecov3}
\begin{eqnarray}
\mathcal{C}_{\eta=\pm1;p_1=p_2=1}(\hat{E}_1,\hat{E}_2)=\pm\frac{4\pi\hbar\omega}{V}\cos(\vartheta_1-\vartheta_2)\\
\mathcal{C}_{\eta=\pm1;p_1=1,p_2=0}(\hat{E}_1,\hat{E}_2)=\mp\frac{4\pi\hbar\omega}{V}\cos(\vartheta_1+\vartheta_2).
\end{eqnarray}
\end{subequations}
From Eqs.~(\ref{Ecov3}) it results that the electric field
covariances for the maximally entangled number states of
Eqs.~(\ref{Ent01}), differs from zero and they have the same
absolute value of that one of Eq.~(\ref{Ecov2}), relating to the
entangled GBSs of Eq.~(\ref{Entbinom1}).

So, the electric fields in two separate cavities prepared in an
entangled state of the form given in Eqs.~(\ref{Entbinom1}),
(\ref{Ent01}), are correlated. However, since the cavity electric
fields are not easily measurable \cite{har}, we cannot acquire any
direct information about non-locality by the electric field
correlations. Thus, it appears to be necessary introducing a
measurable cavity operator to test non-locality of entangled
cavity fields.

\section{\label{bellviol}Bell's inequality}
In view of the previous considerations, in this section we shall
approach the problem of non-locality for entangled fields in two
spatially separate cavities, by looking if it is possible to test
the property of non-locality directly for the entangled cavity
field state of Eq.~(\ref{Entbinom1}). For this purpose, we shall
utilize the Clauser-Horne-Shimony-Holt (CHSH) form of Bell's
inequality \cite{clau,mand} which states that according to any
local hidden variable theory the correlations of two dichotomic
observables $A_1(a_1),A_2(a_2)$ relative to two correlated
subsystems $1,2$, characterized by the parameters $a_1,a_2$ and
whose measurement can have only two possible outcomes labelled
$\pm1$, must satisfy the following inequality
\begin{eqnarray}
S_B&\equiv&\modul{\ave{A_1(a_1)A_2(a_2)}-\ave{A_1(a_1)A_2(a'_2)}}\nonumber\\
&+&\modul{\ave{A_1(a'_1)A_2(a_2)}+\ave{A_1(a'_1)A_2(a'_2)}}\leq2\label{BellIneq}
\end{eqnarray}
where $S_B$ is called \emph{Bell's function}.

\subsection{\label{dicop}Dichotomic field operator}
To test the CHSH form of Bell's inequality defined in
Eq.~(\ref{BellIneq}) for the cavity field entanglement of
Eq.~(\ref{Entbinom1}), one must choose the appropriate operator
which correspond to the observable to be measured.

Let us first consider one single-mode cavity. Utilizing the GBS
$\ket{p,\phi}$ and the orthogonal GBS $\ket{1-p,\pi+\phi}$, we
introduce the dichotomic operator $\hat{F_p}(\phi)$, acting on the
field mode and characterized by the field parameters $p,\phi$,
defined as
\begin{equation}
\hat{F_p}(\phi)\equiv\ket{p,\phi}\bra{p,\phi}-\ket{1-p,\pi+\phi}\bra{1-p,\pi+\phi}.\label{F}
\end{equation}
This operator has eigenvalues $\pm1$ for any values of the
parameters $p,\phi$ and its corresponding expression in the Fock
space basis is
\begin{eqnarray}
\hat{F_p}(\phi)&=&(2p-1)\left(\ket{1}\bra{1}-\ket{0}\bra{0}\right)\nonumber\\
&+&2\sqrt{p(1-p)}\left(e^{i\phi}\ket{1}\bra{0}+e^{-i\phi}\ket{0}\bra{1}\right).\label{FFock}
\end{eqnarray}
We shall test Bell's inequality of Eq.~(\ref{BellIneq}) using the
operators $\hat{F_p}(\phi)$, for each cavity, with different
phases $\phi$ but with the same $p$. The matrix representation of
$\hat{F_p}(\phi')$, in the basis
$\mathcal{B}_{p,\phi}=\{\ket{p,\phi},\ket{1-p,\pi+\phi}\}$ of two
orthogonal GBSs, is
\begin{equation}
\hat{F_p}(\phi')=\left(\begin{array}{lr}F_{11}&F_{12}\\F_{12}^{\ast}&-F_{11}\end{array}\right),\label{matrixF}
\end{equation}
where, using Eqs.~(\ref{FFock}) and (\ref{bin1}), the matrix
elements $F_{11},F_{12}$ have the explicit expressions
\begin{eqnarray}
F_{11}&=&\bra{p,\phi}\hat{F_p}(\phi')\ket{p,\phi}\nonumber\\
&=&1-8p(1-p)\sin^{2}\big((\phi'-\phi)/2\big);\nonumber\\
F_{12}&=&\bra{p,\phi}\hat{F_p}(\phi')\ket{1-p,\pi+\phi}\nonumber\\
&=&2\sqrt{p(1-p)}\big[2(1-2p)\sin^{2}\big((\phi'-\phi)/2\big)\nonumber\\&&\hspace{1.85cm}+\
i\sin(\phi'-\phi)\big]. \label{elemF}
\end{eqnarray}
The matrix of Eq.~(\ref{matrixF}) has clearly eigenvalues $\pm1$,
in fact
\begin{equation}
\lambda^{2}=F_{11}^{2}+\modul{F_{12}}^{2}=1\Rightarrow\lambda=\pm1.
\end{equation}
The corresponding eigenstates $\ket{p,\phi'},\ket{1-p,\pi+\phi'}$
of the operator $\hat{F_p}(\phi')$ of Eq.~(\ref{matrixF}),
expressed in the basis $\mathcal{B}_{p,\phi}$, result to be
\begin{eqnarray}
\ket{1,p,\phi'}&=&\frac{1}{\mathcal{N}_{F}}\bigg[\modul{F_{12}}\ket{1,p,\phi}\nonumber\\
&+&\frac{(1-F_{11})\modul{F_{12}}}{F_{12}}\ket{1,1-p,\pi+\phi}\bigg],\nonumber\\
\ket{1,1-p,\pi+\phi'}&=&\frac{1}{\mathcal{N}_{F}}\bigg[\frac{(F_{11}-1)\modul{F_{12}}}{F_{12}^{\ast}}\ket{1,p,\phi}\nonumber\\
&+&\modul{F_{12}}\ket{1,1-p,\pi+\phi}\bigg], \label{Auto3}
\end{eqnarray}
where the normalization constant $\mathcal{N}_{F}$ is given by
\begin{equation}
\mathcal{N}_{F}=\sqrt{\modul{F_{12}}^{2}+(1-F_{11})^{2}}.
\end{equation}

In subsection~\ref{exptest} we will show that the operator
$\hat{F_p}(\phi)$ defined in Eq.~(\ref{F}), with $p$ fixed and for
different values of $\phi$, is an observable, i.e. its eigenvalues
are experimentally measurable.

\subsection{\label{belltest}Bell's inequality violation}
Let us consider the entangled GBSs introduced in
Eq.~(\ref{Entbinom1}) and for the sake of simplicity set
\begin{equation}
p_1=p_2=p,\quad\vartheta_1=\vartheta_2=\vartheta.\label{pvar}
\end{equation}
With this choice, the entangled state of Eq.~(\ref{Entbinom1})
takes the form
\begin{eqnarray}
\ket{\Psi}&=&\mathcal{N_{\eta}}\big[\ket{p,\vartheta}_1\ket{1-p,\pi+\vartheta}_2\nonumber\\
&+&\eta\ket{1-p,\pi+\vartheta}_1\ket{p,\vartheta}_2\big]\label{Entbin1}
\end{eqnarray}\\ ${}$\\
where $\eta$ is real and ${\cal
N}_\eta=1/\sqrt{1+\modul{\eta}^{2}}$. The choice of
Eq.~(\ref{pvar}) allows us to simplify the expressions, so to have
the results in a more readable form but without loss of
generality.

Substituting the observables $A_j(a_j)$ of Eq.~(\ref{BellIneq})
with the dichotomic operators $\hat{F}_p^{(j)}(\phi_j)$ ($j=1,2$
now indicates the cavity~$j$) defined in Eq.~(\ref{F}), with a
value of the probability of single photon occurrence $p$ fixed and
equal to that one appearing in the entangled state given in
Eq.~(\ref{Entbin1}), Bell's inequality of Eq.~(\ref{BellIneq}) can
be written as
\begin{eqnarray}
S_B&=&\modul{\ave{\hat{F}_p^{(1)}(\phi_1)\hat{F}_p^{(2)}(\phi_2)}-\ave{\hat{F}_p^{(1)}(\phi_1)\hat{F}_p^{(2)}(\phi'_2)}}\nonumber\\
&+&\modul{\ave{\hat{F}_p^{(1)}(\phi'_1)\hat{F}_p^{(2)}(\phi_2)}+\ave{\hat{F}_p^{(1)}(\phi'_1)\hat{F}_p^{(2)}(\phi'_2)}}\leq2.\nonumber
\\\label{BellF}
\end{eqnarray}
The quantum correlations appearing in Eq.~(\ref{BellF}) are given
by
\begin{equation}
\ave{\hat{F}_p^{(1)}(\phi_1)\hat{F}_p^{(2)}(\phi_2)}=\bra{\Psi}\hat{F}_p^{(1)}(\phi_1)\hat{F}_p^{(2)}(\phi_2)\ket{\Psi}\label{corrFdef}
\end{equation}
where $\ket{\Psi}$ is the entanglement of GBSs of
Eq.~(\ref{Entbin1}).

Using Eqs.~(\ref{elemF}), (\ref{Entbin1}) and (\ref{corrFdef}),
the correlation function
$\ave{\hat{F}_p^{(1)}(\phi_1)\hat{F}_p^{(2)}(\phi_2)}$ is given by
\begin{widetext}
\begin{eqnarray}
\ave{\hat{F}_p^{(1)}(\phi_1)\hat{F}_p^{(2)}(\phi_2)}&=&-1+8p(1-p)\bigg\{\sin^2\frac{\phi_1-\vartheta}{2}+\sin^2\frac{\phi_2-\vartheta}{2}
-8p(1-p)\sin^2\frac{\phi_1-\vartheta}{2}\sin^2\frac{\phi_2-\vartheta}{2}\nonumber\\
&+&\frac{\eta}{1+\modul{\eta}^2}\bigg[4(1-2p)^{2}\sin^2\frac{\phi_1-\vartheta}{2}\sin^2\frac{\phi_2-\vartheta}{2}
+\sin(\phi_1-\vartheta)\sin(\phi_2-\vartheta)\bigg]\bigg\}.\label{CorrF}
\end{eqnarray}
\end{widetext}
It depends on the probability of single photon occurrence $p$, on
the phase angle $\vartheta$ characteristic of the GBSs of the
entanglement $\ket{\Psi}$, on the phase angles $\phi_j$ of the
cavity operators $\hat{F}_p^{(j)}(\phi_j)$ and on the parameter of
entanglement $\eta$. We also note that all arguments of the
trigonometric functions of Eq.~(\ref{CorrF}) are shifted of the
same angle $-\vartheta$, so the angle $\vartheta$ can be
arbitrarily fixed. Now we shall look for a set of values of the
parameters $p,\phi_1,\phi_2,\phi'_1,\phi'_2,\eta$ such that a
violation of Bell's inequality of Eq.~(\ref{BellF}) occurs, i.e.
$S_B>2$.

We begin by looking for the value of $p$ which maximizes Bell's
function $S_B$. As $S_B$ is formed by correlations of the kind
(\ref{CorrF}), all having the same dependence on $p$, we simply
set the partial derivative relating to $p$ of the correlation
functions appearing in Eq.~(\ref{BellF}) equal to zero. We have
\begin{equation}
\frac{\partial S_B}{\partial
p}=(1-2p)f(p,\phi_{1},\phi_{2},\phi'_1,\phi'_2,\eta)=0,\label{p}
\end{equation}
where $f(p,\phi_{1},\phi_{2},\phi'_1,\phi'_2,\eta)$ is a
non-singular function of the system variables. So, from
Eq.~(\ref{p}), we obtain
\begin{equation}
p=1/2.
\end{equation}
It is possible to see that this value of $p$ corresponds to a
maximum of Bell's function $S_B$. For this value $p=1/2$, setting
$\hat{F}_{1/2}^{(j)}(\phi_j)\equiv\hat{F}^{(j)}(\phi_j)$ for
simplicity, the correlation function of Eq.~(\ref{CorrF}) becomes
\begin{eqnarray}
\ave{\hat{F}^{(1)}(\phi_1)\hat{F}^{(2)}(\phi_2)}&=&\frac{2\eta}{1+\modul{\eta}^2}\sin(\phi_1-\vartheta)\sin(\phi_2-\vartheta)\nonumber\\
&-&\cos(\phi_1-\vartheta)\cos(\phi_2-\vartheta)\label{CorrF1}
\end{eqnarray}
At this point, it is useful to consider the \emph{degree of
entanglement} of the state $\ket{\Psi}$ of Eq.~(\ref{Entbin1}),
defined as \cite{abour}
\begin{equation}
G^{(E)}=G^{(E)}(\modul{\eta})\equiv\frac{2\modul{\eta}}{1+\modul{\eta}^{2}}.\label{entdegree}
\end{equation}
$G^{(E)}$ is invariant with respect to the substitution
$\modul{\eta}\rightarrow1/\modul{\eta}$, equal to zero for
$\modul{\eta}=0,+\infty$ and equal to one (maximum value) for
$\modul{\eta}=1$. Using the expression of the correlation function
given in Eq.~(\ref{CorrF1}) and the definition of $G^{(E)}$ of
Eq.~(\ref{entdegree}), Bell's function of Eq.~(\ref{BellF}) can be
written as
\begin{eqnarray}
S_B&=&\big|G^{(E)}\sin(\phi_1-\vartheta)[\sin(\phi_2-\vartheta)-\sin(\phi'_2-\vartheta)]\nonumber\\
&\pm&\cos(\phi_1-\vartheta)[\cos(\phi_2-\vartheta)-\cos(\phi'_2-\vartheta)]\big|\nonumber\\
&+&\big|G^{(E)}\sin(\phi'_1-\vartheta)[\sin(\phi_2-\vartheta)+\sin(\phi'_2-\vartheta)]\nonumber\\
&\pm&\cos(\phi'_1-\vartheta)[\cos(\phi_2-\vartheta)+\cos(\phi'_2-\vartheta)]\big|,
\label{SBell}
\end{eqnarray}
where the signs $\pm$ correspond to $\eta$ negative or positive,
respectively. We show that, with $p=1/2$ and appropriate choices
of the phase angles $\phi_{1},\phi_{2},\phi'_{1},\phi'_{2}$,
Bell's function of Eq.~(\ref{SBell}) is greater than two, $S_B>2$,
and Bell's inequality of Eq.~(\ref{BellF}) is violated for a wide
range of the degree of entanglement $G^{(E)}$. We give here the
particular case where the largest possible quantum mechanical
violation of Bell's inequality occurs (maximal violation). In
App.\ref{violapp} we give another interesting particular case.

\subsubsection{Maximal violation}
\begin{figure}
\includegraphics[width=0.4\textwidth, height=0.2\textheight]{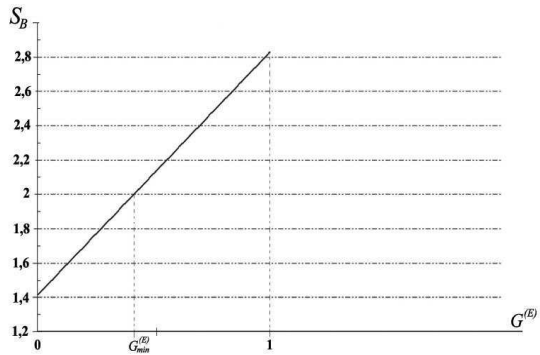}
\caption{\label{grafBellmax}\textbf{Maximal violation}. Bell's
function $S_B$ \emph{vs} $G^{(E)}$ with the conditions of
Eq.~(\ref{condmax}) and $p=1/2$. $G_{min}^{(E)}=\sqrt{2}-1$. For
$G^{(E)}=1$ we obtain the maximal violation
$S_B^{max}=2\sqrt{2}$.}
\end{figure}
We choose the following values of the phase angles
\begin{eqnarray}
\phi_1&=&\vartheta,\quad \phi_2=\vartheta+\pi/4,\nonumber\\
\phi'_1&=&\vartheta+\pi/2,\quad
\phi'_2=\vartheta+3\pi/4.\label{condmax}
\end{eqnarray}
This is the standard choice of the angles for spin-$1/2$ objects
to obtain the maximal violation of Bell's inequality of the kind
(\ref{BellIneq}) \cite{pre,gerry,mand}, where the separation
between two consecutive angles is $\pi/4$.

Substituting the values of the angles of Eq.~(\ref{condmax}) in
Eq.~(\ref{SBell}), Bell's function $S_B$ results to be
\begin{equation}
S_B=\sqrt{2}(1+G^{(E)}).\label{SumBell3}
\end{equation}
Using Eq.~(\ref{SumBell3}), we obtain that Bell's inequality given
in Eq.~(\ref{BellF}) is violated, i.e. $S_B>2$, for those values
of the degree of entanglement $G^{(E)}$ inside the interval
\begin{equation}
\textrm{$G_{min}^{(E)}<G^{(E)}\leq1$, with
$G_{min}^{(E)}=\sqrt{2}-1\approx0,41$}
\end{equation}
where $G_{min}^{(E)}$ is obtained by $S_B=2$. The graph of Bell's
function $S_B$ of Eq.~(\ref{SumBell3}) is plotted in
Fig.\ref{grafBellmax}. For $G^{(E)}=1$, Bell's function $S_B$ of
Eq.~(\ref{SumBell3}) takes its maximum value
\begin{equation}
S_B(G^{(E)}=1)=S_B^{max}=2\sqrt{2}\approx2,8284.
\end{equation}
It is possible to prove \cite{pre} that this is the maximal
violation of Bell's inequality.

\subsection{\label{exptest}Test of Bell's inequality}
In this section we propose a simple scheme for testing Bell's
inequality violation shown in subsection~\ref{belltest}. This test
gives, in principle, a direct experimental demonstration of
non-locality for the entangled GBSs in two separate cavities given
in Eq.~(\ref{Entbin1}). We show that this test permits to measure
in a simple way, for a single cavity, the eigenvalues ($\pm1$) of
the cavity operator $\hat{F}_p(\phi)$, by a resonant two-level
probe atom which ``reads'' the cavity field. This result is
obtained by associating the outcome of the final atomic state
measurement with a given eigenvalue of the operator
$\hat{F}_p(\phi)$. We first describe the dynamics of the resonant
atom-cavity interaction.

Let $\ket{\uparrow}$ and $\ket{\downarrow}$ be respectively the
excited and ground state of the probe atom with transition
frequency resonant with the cavity field mode $\omega$. The
dynamics of the resonant atom-cavity interaction is governed by
the usual Jaynes-Cummings Hamiltonian \cite{jay}
\begin{equation}
H_{J-C}=\frac{1}{2}\hbar\omega\sigma_{z}+\hbar\omega
a^{\dag}a+i\hbar g(\sigma_{+}a-\sigma_{-}a^{\dag}),\label{H}
\end{equation}
where $g$ is the atom-field coupling constant, $a$ and $a^{\dag}$
are the field annihilation and creation operators and
$\sigma_{z}$, $\sigma_{+}$, $\sigma_{-}$ are the pseudo-spin
atomic operators
\[\sigma_z=\ket{\uparrow}\bra{\uparrow}-\ket{\downarrow}\bra{\downarrow},\quad\sigma_+=\ket{\uparrow}\bra{\downarrow},\quad
\sigma_-=\ket{\downarrow}\bra{\uparrow}.\nonumber\] The
Jaynes-Cummings Hamiltonian of Eq.~(\ref{H}) generates the
transitions \cite{mey,compagno}
\begin{eqnarray}
\ket{\uparrow n}&\rightarrow&\cos(g\sqrt{n+1}t)\ket{\uparrow n}-\sin(g\sqrt{n+1}t)\ket{\downarrow n+1}\nonumber\\
\ket{\downarrow n}&\rightarrow&\cos(g\sqrt{n}t)\ket{\downarrow
n}+\sin(g\sqrt{n}t)\ket{\uparrow n-1},\label{evo}
\end{eqnarray}
where $n$ is the number of photons inside the cavity. We ignore
atomic and field dissipations during the atom-field interaction,
which is a good approximation for such a systems constituted by
Rydberg atoms and high-$Q$ cavities \cite{har}.

\begin{figure}
\includegraphics[width=0.46\textwidth, height=0.13\textheight]{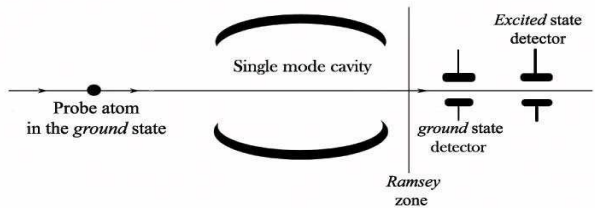}
\caption{\label{fig2probeatom} Experimental scheme for measuring
the dichotomic field operator $\hat{F}_p(\phi)$ in a cavity. The
atom-cavity interaction time is $T_P=\pi/2g$, where $g$ is the
coupling constant.}
\end{figure}

The experimental measurement scheme is shown in
Fig.\ref{fig2probeatom}.
A two-level probe atom is initially prepared in the ground state
$\ket{\downarrow}$ and interacts on resonance with the cavity
field for a time given by
\begin{equation}
T_P=\pi/2g,
\end{equation}
where $g$ is the atom-field coupling constant. At this point of
the sequence, utilizing the Jaynes-Cummings evolutions
(\ref{evo}), we find the following atom-field transitions
\begin{eqnarray}
\ket{\downarrow}\ket{p,\phi}&\stackrel{T_P}{\rightarrow}&\ket{0}(\sqrt{1-p}\ket{\downarrow}+e^{i\phi}\sqrt{p}\ket{\uparrow});\nonumber\\
\ket{\downarrow}\ket{1-p,\pi+\phi}&\stackrel{T_P}{\rightarrow}&\ket{0}(\sqrt{p}\ket{\downarrow}-e^{i\phi}\sqrt{1-p}\ket{\uparrow})
\label{probevo1}
\end{eqnarray}
where $\ket{p,\phi},\ket{1-p,\pi+\phi}$ are the eigenstates of the
operator $\hat{F}_p(\phi)$ (see Eq.~(\ref{F})). Observing
Eqs.~(\ref{probevo1}), we note that a measurement of the atomic
state after the interaction with the cavity does not allow us to
distinguish the two initial eigenstates of $\hat{F}_p(\phi)$.

To obtain this, after going out of the cavity, we let the atom
cross an opportunely set Ramsey zone where it, interacting with a
classical microwave field, undergoes the transformations
\begin{eqnarray}
\ket{\uparrow}&\rightarrow&\ket{\uparrow}_{\boldsymbol{u}}\equiv\cos(\theta/2)\ket{\uparrow}-e^{i\varphi}\sin(\theta/2)\ket{\downarrow}\nonumber\\
\ket{\downarrow}&\rightarrow&\ket{\downarrow}_{\boldsymbol{u}}\equiv
e^{-i\varphi}\sin(\theta/2)\ket{\uparrow}+\cos(\theta/2)\ket{\downarrow}
\label{ramsequ}
\end{eqnarray}
where the versor $\bm{u}$ is
\begin{equation}
\bm{u}\equiv(-\sin\theta\cos\varphi,-\sin\theta\sin\varphi,\cos\theta).\label{versoubis}
\end{equation}
The angle $\theta\in]0,\pi[$ is the so-called ``Ramsey pulse''.
The values of $\theta,\varphi$ are fixed by adjusting the
classical Ramsey field amplitude and the interaction time so to
have
\begin{equation}
\cos(\theta/2)=\sqrt{p},\
\sin(\theta/2)=\sqrt{1-p};\quad\varphi=-\phi\label{tetaramsey}
\end{equation}
where the values of $p,\phi$ are equal to the ones of the operator
$\hat{F}_p(\phi)$ to be measured. Using Eqs.~(\ref{ramsequ})
together with Eqs.~(\ref{tetaramsey}), after the Ramsey zone
interaction, we find that the total atom-cavity states of
Eqs.~(\ref{probevo1}) undergo the following evolutions
\begin{subequations}
\label{probevo2}
\begin{eqnarray}
\ket{\downarrow}\ket{p,\phi}&\rightarrow&\ket{0}e^{i\phi}\ket{\uparrow}\\
\ket{\downarrow}\ket{1-p,\pi+\phi}&\rightarrow&\ket{0}\ket{\downarrow}.
\end{eqnarray}
\end{subequations}
At the end of the experimental sequence of Fig.\ref{fig2probeatom}
the atomic state is measured by field ionization detectors. From
the final atom-field states of Eqs.~(\ref{probevo2}) and from the
definition of $\hat{F}_p(\phi)$ given in Eq.~(\ref{F}), we
immediately obtain that:
\begin{itemize}
\item the measurement of the excited atomic state $\ket{\uparrow}$
corresponds to the eigenvalue $+1$ of $\hat{F}_p(\phi)$;

\item the measurement of the ground atomic state
$\ket{\downarrow}$ corresponds to the eigenvalue $-1$ of
$\hat{F}_p(\phi)$.
\end{itemize}
It must be noted from Eqs.~(\ref{probevo2}) that, at the end of
the sequence, the cavity field state is always in the vacuum
state.

Utilizing these results and the ones of subsection~\ref{belltest},
we see that an experimental test of the CHSH form of Bell's
inequality of Eq.~(\ref{BellF}) for the entangled GBSs in two
spatially separate cavities of Eq.~(\ref{Entbin1}) requires the
following steps:
\begin{itemize}
\item[i)] the generation of the entangled GBSs given in
Eq.~(\ref{Entbin1}), with $p=1/2$ and an arbitrarily fixed
$\vartheta$;

\item[ii)] the resonant interaction of two two-level probe atoms
with their respective cavity and with a successive Ramsey zone,
according to the scheme of Fig.\ref{fig2probeatom}. The Ramsey
zone interaction is set so as to have a $\pi/2$-pulse and the
values of the angles $\phi_j$ desired (see
Eq.~(\ref{tetaramsey})), given, for example, by the choice of
Eq.~(\ref{condmax});

\item[iii)] the simultaneous measurement of atomic states, each
one corresponding to a given eigenvalue of the dichotomic operator
$\hat{F}_{1/2}^{(j)}(\phi_j)$, as it was found in
Eqs.~(\ref{probevo2}). The simultaneity of the measurements
ensures that the events are ``space-like'' separate, so as to
close the locality loophole \cite{massar}.
\end{itemize}
After repeating this sequence for many times, it is possible to
obtain the correlations
$\ave{\hat{F}_{1/2}^{(1)}(\phi_{1})\hat{F}_{1/2}^{(2)}(\phi_{2})}$
for the desired values of the angles $\phi_1,\phi_2$ by
statistical averages and then to test Bell's inequality of
Eq.~(\ref{BellF}). A brief discussion on the typical experimental
parameters involved in this measurement scheme is given in
Sec.\ref{conc}.

\section{\label{gensch}Sketch of a scheme to generate entangled generalized Bernoulli states in separate cavities}
\begin{figure}
\includegraphics[width=0.46\textwidth, height=0.15\textheight]{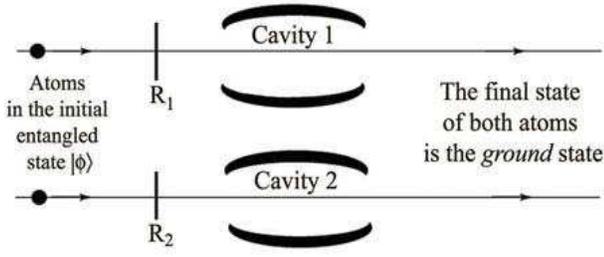}
\caption{\label{entfig1} Experimental scheme for the generation of
entangled orthogonal GBSs in two separate cavities.}
\end{figure}

As stressed at the end of the previous section, testing Bell's
inequality violation requires, as first step, the generation of
entangled GBSs of Eq.~(\ref{Entbinom1}) in the two separate
single-mode cavities. In this section, following standard
procedures currently implemented in laboratory to produce assigned
states of the electromagnetic fields inside a cavity
\cite{har,har1}, we sketch the main steps of a scheme, shown in
Fig.\ref{entfig1}, aimed at generating in principle our target
entangled state.

Let us consider a couple of identical two-level Rydberg atoms
initially in the entangled state
\begin{equation}
\ket{\phi}={\cal
N}_\eta(\ket{\uparrow_1\downarrow_2}+\eta\ket{\downarrow_1\uparrow_2}),\label{atent12}
\end{equation}
where $\eta$ is real and ${\cal
N}_\eta=1/\sqrt{1+\modul{\eta}^{2}}$ is a normalization constant,
which can be prepared using, for example, the scheme suggested by
Gerry \cite{gerry} and let us also consider two identical high-$Q$
single-mode cavities each in the vacuum state $\ket{O_1}$,
$\ket{0_2}$. Each atom of the entangled atomic state of
Eq.~(\ref{atent12}) crosses a Ramsey zone ${\rm R}_j$ ($j=1,2$),
where it undergoes the transformations given in
Eqs.~(\ref{ramsequ}). If we set, for each atom~$j$,
\begin{equation}
\cos(\theta_j/2)\equiv\sqrt{p_j},\
\sin(\theta_j/2)\equiv\sqrt{1-p_j};\
\varphi_j=-\vartheta_j\label{cosp}
\end{equation}
where $p_1,p_2$ are two arbitrary real numbers inside the interval
$]0,1[$, the transformations of Eqs.~(\ref{ramsequ}) can be
written as
\begin{eqnarray}
\ket{\uparrow_j}&\rightarrow
&\ket{\uparrow}_{\boldsymbol{u}_j}\equiv\sqrt{p_j}\ket{\uparrow_j}-e^{-i\vartheta_j}\sqrt{1-p_j}\ket{\downarrow_j}\nonumber\\
\ket{\downarrow_j}&\rightarrow
&\ket{\downarrow}_{\boldsymbol{u}_j}\equiv
e^{i\vartheta_j}\sqrt{1-p_j}\ket{\uparrow_j}+\sqrt{p_j}\ket{\downarrow_j}.\label{ramseqj}
\end{eqnarray}\\
The total atom-cavity state after the Ramsey zones is
\begin{equation}
\ket{\Psi(0)}_{tot}=\mathcal{N}_{\eta}\big(\ket{\uparrow}_{\bm{u}_1}\ket{\downarrow}_{\bm{u}_2}+
\eta\ket{\downarrow}_{\bm{u}_1}\ket{\uparrow}_{\bm{u}_2}\big)\ket{0_10_2}.\label{Psitotin}
\end{equation}
Now, the atoms resonantly interact with their respective cavities
for a time $T$ such that
\begin{equation}
gT=\pi/2\label{Tempint1}.
\end{equation}
Since the two subsystems 1 and 2 are independent, we can utilize
the usual Jaynes-Cummings evolutions given in Eqs.~(\ref{evo}) for
each subsystem $j$, so that at the time $T$ the state of
Eq.~(\ref{Psitotin}) becomes
\begin{eqnarray}
\ket{\Psi(T)}_{tot}&=&\mathcal{N_{\eta}}\Big[\ket{p_1,\vartheta_1}_1\ket{1-p_2,\pi+\vartheta_2}_2\nonumber\\
&+&\eta\ket{1-p_1,\pi+\vartheta_1}_1\ket{p_2,\vartheta_2}_2\Big]\ket{\downarrow_1\downarrow_2},\nonumber\\\label{Psitotal}
\end{eqnarray}
where we have omitted the unimportant global phase factor
$e^{i(\pi-\vartheta_1-\vartheta_2)}$ and we have used the notation
of Eq.~(\ref{bin1}) for the GBSs. After the interaction, the total
atom-cavity state of Eq.~(\ref{Psitotal}) describes both atoms in
their ground state and the cavity field state in the pure
entangled state
\begin{eqnarray}
\ket{\Psi}&=&\mathcal{N_{\eta}}\Big[\ket{p_1,\vartheta_1}_1\ket{1-p_2,\pi+\vartheta_2}_2\nonumber\\
&+&\eta\ket{1-p_1,\pi+\vartheta_1}_1\ket{p_2,\vartheta_2}_2\Big]\label{Entbinom}
\end{eqnarray}
which represents the entanglement of two orthogonal GBSs of the
form defined in Eq.~(\ref{Entbinom1}). By adjusting appropriately
the settings of the Ramsey zones, the values of $p_j$ and
$\vartheta_j$ can be arbitrarily changed (see Eq.~(\ref{cosp})).

Although we shall not enter into the details of the experimental
feasibility of the proposed generation scheme, we shall give here
a brief valuation of some potential errors involved in such a
scheme. A necessary condition required by our generation scheme is
that the atoms interact with the cavities for a given period of
time. This can be obtained by the selection of a well determinate
atomic velocity. Any experimental error $\Delta v$ in the atomic
velocity $v$ induces an error $\Delta T$ in the atom-cavity
interaction time $T$ given by
\begin{equation}
\Delta
T=\bigg|L\Delta\bigg(\frac{1}{v}\bigg)\bigg|\approx\frac{L\Delta
v}{v^2}=T\frac{\Delta v}{v},\label{timerror}
\end{equation}
where $L$ is the cavity length. From Eq.~(\ref{timerror}) we find
that the velocity relative error $\Delta v/v$ must satisfy the
condition
\begin{equation}
\frac{\Delta T}{T}\approx\frac{\Delta v}{v}\ll1\label{velerr}
\end{equation}
in order that the time relative error $\Delta T/T$ may be
negligible. In current laboratory experiments it is possible to
select a given atomic velocity with a relative error $\sim10^{-2}$
or less \cite{hag,har1}, so that, from Eq.~(\ref{velerr}), $\Delta
T/T$ is of the same order.

Another aspect we have excluded is the atomic or photon decay
during the atom-cavity interactions. This assumption can be held
if
\begin{equation}
\tau_{at},\tau_{cav}>T,\label{condat}
\end{equation}
where $\tau_{at},\tau_{cav}$ are respectively the atomic and
photon mean lifetimes and $T$ is the interaction time. For Rydberg
atomic levels and microwave superconducting cavities with quality
factor $Q\sim10^8\div10^{10}$, we have
$\tau_{at}\sim10^{-5}\div10^{-2}\textrm{s}$ and
$\tau_{cav}\sim10^{-4}\div10^{-1}\textrm{s}$. Since typical
atom-cavity field interaction times are
$T\sim10^{-5}\div10^{-4}\textrm{s}$, the required condition of
Eq.~(\ref{condat}) can be satisfied \cite{har}. Moreover, the
typical mean lifetimes of the Rydberg atomic levels $\tau_{at}$
must be such that the atoms do not decay during the entire
sequence of the scheme and the photon mean lifetimes $\tau_{cav}$
must be long enough to permit cavity fields not to decay before
they interact with probe atoms \cite{har,har1,nogu}, so as to
allow the successive Bell's inequality test (see
subsection~\ref{exptest}).

\section{\label{conc}Conclusion}
In this paper, we have considered two spatially separate
single-mode cavities where it has been injected an electromagnetic
field in an entangled state of two orthogonal generalized
Bernoulli states (GBS). We have then studied the expectation
values and the correlations of the electric fields of these two
cavities, finding that they are in general different from zero.
The existence of such correlations between the two cavities
indicates that the fields are non-local in this system. The cavity
electric fields, and so their correlations, are however not
directly measurable.

To examine the non-local feature of the electromagnetic fields in
the two cavities, we have introduced, for each cavity, a
dichotomic operator $\hat{F_p}(\phi)$ with eigenvalues $\pm1$
acting on the field states and in principle measurable. Using the
quantum correlations of couples of these operators with different
values of $\phi$ and with the same $p$, we have constructed the
CHSH form of Bell's inequality finding that, for opportune choices
of the system variables, this Bell's inequality is violated for a
wide range of the degree of entanglement. Quantum non-locality is
thus directly shown by the cavity fields in the entanglement of
GBSs in two separate cavities. This remarkable feature of such an
entangled state is one of our main results.

We have also proposed a simple test of this Bell's inequality
violation which exploits a couple of two-level probe atoms each
interacting resonantly with the respective cavity and successively
with an opportune Ramsey zone for an assigned time. Appropriate
Ramsey zone settings, i.e. pulse and relative phase of the atomic
states, allow the unconditional measurement of the cavity operator
$\hat{F_p}(\phi)$. Our result is that, if each probe atom is
initially in the ground state and its final state is detected at
the end of the sequence, the measurement of the excited or ground
state is equivalent to the eigenvalue $+1$ or $-1$ of the cavity
operator, respectively. So, our Bell's inequality test requires
(\textit{i}) the repeated preparation of entangled orthogonal GBSs
in two separate single-mode cavities, (\textit{ii}) the use of two
independent probe atoms, each of them following the above
experimental sequence with the desired settings of Ramsey zone,
and (\textit{iii}) the simultaneous measurement of the final
atomic states, which is equivalent to the measurement of the
eigenvalues of the introduced cavity operators. The cavity
operators correlations can be then obtained by statistical
averages of the eigenvalues products. This Bell's inequality test
is non-conditional and it requires that the atoms resonantly
interact with the cavities for an assigned time. The main result
found in this paper is that, by our procedure, it is possible to
obtain in a simple way a direct verification of Bell's inequality
violation for the cavity electromagnetic fields. We wish to
emphasize that our test is immune to the typical experimental
errors on the desired interaction time \cite{hag,har1}. In the
experimental context of the Bell's inequality test, another
important parameter to be considered is the atomic state detector
efficiency $\alpha$ that we have supposed as ideal for simplicity.
Had we incorporated detectors efficiencies in the correlation
functions, then the CHSH form of Bell's inequality would not have
been violated for values of $\alpha$ less than
$\alpha_t=2/(\sqrt{2}+1)\approx0,8284$ for maximally entangled
states of a bipartite system \cite{clau,massar}. However, also in
this case, it is always possible to test the CHSH form of Bell's
inequality with the ``fair sampling'' hypothesis that the
sub-ensemble of detected events (detected atoms) represents the
whole ensemble. So, the results just rely on the detected events
but the probabilistic nature of atomic detection leaves ``open''
the detection loophole \cite{moeh,massar}. Only for detector
efficiencies greater than $\alpha_t$ the detection loophole can be
closed.

We conclude that, at this time, the experimental developments seem
to be rather promising on the possibility of implementing our
measurement scheme, so as to allow the realization of the first
direct test of Bell's inequality for entangled fields in two
spatially separate cavities.

\appendix
\section{\label{binapp}Binomial states. Definition and some properties}
In Sec.\ref{entsez} we have given the definition of the particular
generalized binomial state $\ket{N,p,\phi}$. In this paper we
consider the particular case of generalized binomial states with
$N=1$, i.e. the so-called ``generalized Bernoulli state'' (GBS)
\cite{sto}, whose explicit expression is
\begin{equation}
\ket{p,\phi}=\sqrt{1-p}\ket{0}+e^{i\phi}\sqrt{p}\ket{1},\label{bin1}
\end{equation}
as it is readily obtained by Eq.~(\ref{bindef}). We now give some
properties of the generalized binomial state $\ket{N,p,\phi}$,
defined in Eq.~(\ref{bindef}), which will be useful in this paper.

i) In the limits $p=0,1$, the binomial state of Eq.~(\ref{bindef})
becomes
\begin{equation}
\lim_{p\rightarrow0}\ket{N,p,\phi}=\ket{0},\quad
\lim_{p\rightarrow1}\ket{N,p,\phi}=\ket{N}.\label{binlim}
\end{equation}

ii) Two binomial states of the kind given in Eq.~(\ref{bindef}),
$\ket{N,p,\phi},\ket{N,p',\phi'}$, with the same maximum number of
photons $N$, are orthogonal if and only if
\begin{equation}
p'=1-p\quad e\quad \phi'=(2m+1)\pi+\phi,\
m\in\mathbb{Z}.\label{ortog}
\end{equation}
As far as we know, this orthogonality property has not been
reported in literature yet, so we give here the proof.

The scalar product of two binomial states $\ket{N,p,\phi}$ and
$\ket{N,p',\phi'}$, of the form defined in Eq.~(\ref{bindef}), is
given by
\begin{eqnarray}
\scal{N,p,\phi}{N,p',\phi'}&=&\sum_{n=0}^N{N\choose
n}(pp')^{n/2}\nonumber\\&&\times[(1-p)(1-p')]^{(N-n)/2}e^{in(\phi'-\phi)}.\nonumber\\\label{scabin}
\end{eqnarray}
Substituting the equalities of Eq.~(\ref{ortog}) in
Eq.~(\ref{scabin}), we obtain
\begin{eqnarray}
\scal{N,p,\phi}{N,1-p,\pi+\phi}&=&[p(1-p)]^{N/2}\sum_{n=0}^{N}{N\choose
n}e^{in\pi}\nonumber\\
&=&[p(1-p)]^{N/2}(e^{i\pi}+1)^{N}\nonumber\\&=&0,
\end{eqnarray}
where we have used the binomial theorem of Newton and the equality
$e^{i\pi}=-1$. This proves that the conditions of
Eq.~(\ref{ortog}) are sufficient. The conditions of
Eq.~(\ref{ortog}) are also necessary. In fact, the scalar product
of Eq.~(\ref{scabin}) can be wrote
\begin{eqnarray}
\scal{N,p,\phi}{N,p',\phi'}&=&\sum_{n=0}^{N}{N\choose
n}\left(e^{i(\phi'-\phi)}\sqrt{pp'}\right)^{n}\nonumber\\&&\times\left(\sqrt{(1-p)(1-p')}\right)^{N-n}.\nonumber
\end{eqnarray}
Utilizing the usual binomial theorem of Newton, we obtain
\begin{eqnarray}
\scal{N,p,\phi}{N,p',\phi'}&=&\Big(e^{i(\phi'-\phi)}\sqrt{pp'}\nonumber\\&+&\sqrt{(1-p)(1-p')}\Big)^N
\end{eqnarray}
and setting this equation equal to zero, it must be
\begin{equation}
e^{i(\phi'-\phi)}\sqrt{pp'}+\sqrt{(1-p)(1-p')}=0.\label{zero}
\end{equation}
Since the square roots are real and non-negative, from
Eq.~(\ref{zero}) we have
\begin{equation}
e^{i(\phi'-\phi)}=-1\Leftrightarrow \phi'-\phi=(2m+1)\pi,\
m\in\mathbb{Z}.\label{orto1}
\end{equation}
So, Eq.~(\ref{zero}) becomes
\begin{eqnarray}
\sqrt{pp'}=\sqrt{(1-p)(1-p')}&\Rightarrow&
pp'=1-p-p'+pp'\nonumber\\&\Rightarrow&p'=1-p\label{orto2}.
\end{eqnarray}
From the results of Eqs.~(\ref{orto1}), (\ref{orto2}), we see that
the orthogonality condition of Eq.~(\ref{ortog}) is proved.

\section{\label{violapp}Another Bell's inequality violation}
We give here another case of violation of the CHSH form of Bell's
inequality given in Eq.~(\ref{BellF}) for the entangled GBSs of
Eq.~(\ref{Entbin1}). In this case, the interval of values of the
degree of entanglement where a Bell's inequality violation occurs
is the widest obtainable.
\begin{figure}
\includegraphics[width=0.4\textwidth, height=0.2\textheight]{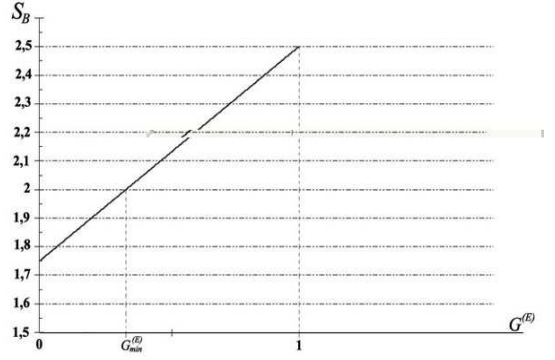}
\caption{\label{grafBell2}Bell's function $S_B$ \emph{vs}
$G^{(E)}$ with the conditions of Eq.~(\ref{Cond2}) together with
$p=1/2$. $G_{min}^{(E)}=1/3$.}
\end{figure}

Let us choose the following values for the phase angles of
Eq.~(\ref{SBell})
\begin{eqnarray}
\phi_1&=&\phi_2=\vartheta,\quad
\phi'_1=\vartheta+\pi/3,\nonumber\\
\phi'_2&=&\left\{\begin{array}{ll}\vartheta-2\pi/3,&\textrm{if
$\eta$ is positive}\\\vartheta+2\pi/3,&\textrm{if $\eta$ is
negative}\end{array}\right.\label{Cond2}
\end{eqnarray}
With this set of values of the angles, Bell's function $S_B$ of
Eq.~(\ref{SBell}) becomes
\begin{equation}
S_B=\frac{7}{4}+\frac{3}{4}G^{(E)}.\label{SumBell2}
\end{equation}
Using Eq.~(\ref{SumBell2}), we find that Bell's inequality of
Eq.~(\ref{BellF}) is violated, i.e. $S_B>2$, for those values of
the degree of entanglement $G^{(E)}$ inside the interval
\begin{equation}
\textrm{$G_{min}^{(E)}<G^{(E)}\leq1$, with
$G_{min}^{(E)}=1/3=0,\bar{3}$.}\label{Gint2}
\end{equation}
The value of $G_{min}^{(E)}$ is clearly obtained by the condition
$S_B=2$. This interval of values of $G^{(E)}$, for which Bell's
inequality is violated, is the widest we found. In
Fig.\ref{grafBell2} we plot the graph of Bell's function $S_B$ of
Eq.~(\ref{SumBell2}) \emph{vs} the degree of entanglement
$G^{(E)}$, for the choices given in Eq.~(\ref{Cond2}). The maximum
violation value of $S_B$ is obtained when $G^{(E)}=1$, i.e. when
the state (\ref{Entbin1}) is maximally entangled, and it is
\begin{equation}
S_B(G^{(E)}=1)=5/2=2,5.
\end{equation}

\end{document}